A 3D Curve Offset Approach for Ruled Surface Generation in Engineering Design

Kun Jiang[1], Xionghui Zhou[1*], Min Li[2], Qiang Niu[1]

1 National Die & Mold CAD Eng. Research Center, Shanghai Jiaotong University, Shanghai, 200030

2 Shanghai Shenmo Die & Mold Manufacturing Co., Ltd., Shanghai, P.R.China, 201203

Abstract: Ruled surface is widely used in engineering design such as parting surface design of injection mold and checking surface design of checking fixture, which are usually generated by offsetting 3D curves. However, in 3D curve offset, there often exist break/interaction/overlapping problems which can't be solved by current CAD software automatically. This paper is targeted at developing a 3D curve offsetting algorithm for ruled surface generation, and three key technologies are introduced in details: An improved curve division method is proposed to reduce the offset accuracy error resulted from different offset distances and curvatures; An offsetting curve overlapping detection and elimination method is proposed; And then, a curve transition method is presented to improve curve offsetting quality for the break and intersection/overlapping regions, where a new algorithm for generating positive weights spherical rational quartic Bezier curve is proposed to bridge the breaks of offset curves to create a smooth ruled surface. Finally, two practical design cases,

[*] Corresponding author.

*Email address*: jiangkacademic@gmail.com (Kun Jiang), xhzhou@sjtu.edu.cn (Xionghui Zhou)

parting surface and checking surface generation, show that the proposed approach can enhance the efficiency and quality for ruled surface generation in engineering design.

## 1 Introduction

Nowadays, there has been a growing interest in ruled surface modeling and machining due to three main reasons (Ding 2009). First, ruled surface-based models are frequently utilized for lots of industrial components, e.g. cams, dies, molds, fan vanes, gas turbine blades, and compressor impellers. Second, free-form surfaces can be approximated using piecewise ruled surfaces. Third, ruled surfaces can be processed in five-axis machining by flank (or side) milling which has many advantages over point milling, such as higher material removal rate, higher machining efficiency and better surface quality.

A ruled surface can be expressed as (Do Carmo 1976):

$$\mathbf{R}(u,v) = \mathbf{P}(u) + v\mathbf{N}(u), (u,v) \in \mathbf{S} \in R^2 \qquad (1)$$

where $\mathbf{S}$ is the (continuous) domain of parameters $u$ and $v$, the curve $\mathbf{P}(u)$ is called the directrix of the surface and a line having $\mathbf{N}(u)$ as direction vector is called a ruling (Ding 2009).

Let $\mathbf{Q}(u) = \mathbf{P}(u) + \mathbf{N}(u)$, where $\mathbf{Q}(u)$ is an 3D offset curve of $\mathbf{P}(u)$ along the vector $\mathbf{N}(u)$, then the formula can be reformed as (Zhang et al. 2010):

$$\mathbf{R}(u,v) = (1-v)\mathbf{P}(u) + v\mathbf{Q}(u), (u,v) \in \mathbf{S} \in R^2 \qquad (2)$$

Hence, the ruled surface can be regarded as the linear interpolation between the original curve and the offset one.

Most researches about curve offsetting focus on planar and geodesic curves, such as tool path generation. Given a parametric curve $\mathbf{C}(t), t \in [0,1]$, its offset curve with an offset distance $d$ is defined by $\mathbf{C}^o(t) = \mathbf{C}(t) + d\mathbf{N}(t)$, where $\mathbf{N}(t)$ is the offset rule which decides the form of the offset curve. For a 2D parameterized curve $\mathbf{C} = (x(t), y(t)), t \in [0,1]$, $\mathbf{N}(t) = (y(t) - x(t))/\sqrt{y^2(t) + x^2(t)}$ is the unitary normal of the original curve $\mathbf{C}(t)$. In general, the offset curve cannot be represented in a polynomial or rational form because of the square root term in the denominator of the unit normal $\mathbf{N}(t)$, so that, it is difficult to obtain the simple forms of these offsets, hence, the approximation methods are needed.

Different from 2D case, the offset rule of a 3D curve is generally not the unit normal of the original curve. In engineering design, $\mathbf{N}(t) = \mathbf{v}(t) \times \mathbf{k}(t)$, where $\mathbf{v}(t)$ is the unit tangent vector of $\mathbf{C}(t)$ (Shin 2003). In parting surfaces design of a mold, $\mathbf{k}(t)$ is the parting direction without relationship to the curve parameter. In the checking surface design of a checking fixture, $\mathbf{k}(t)$ is the unit normal vector of a point on the checked surface. As another difference, the overlapping of 3D offset curves is undesirable in engineering design, since it makes some parts of the ruled surface invisible. However, 3D curve offset tools are not afforded by current commercial CAD software, and designers have to deal with the work manually.

In engineering design, there are many $C^0/G^0$ continuous curves. Actually, the connecting point with the $C^0/G^0$ continuity has two different offset directions, which gives rise to the offsetting curve breaks or intersections. As shown in Figure 1, $Q_1$ is a co-vertex of two edges which belongs to two different surfaces, $Q_2$ is a corner of the parametric region of the surface, $Q_3$ and $Q_4$ are the intersection points of two boundary edges on the surface, with the difference that $Q_3$ is on a convex edge that results in breaks and $Q_4$ on a concave edge that results in intersections.

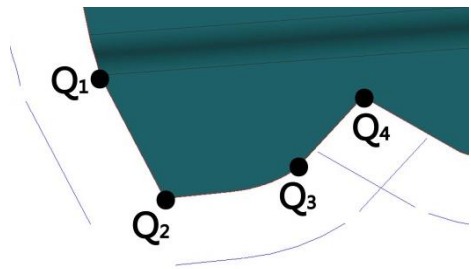

Figure 1. The result of curve offsetting with first-order-discontinuity points

This paper focuses on 3D offsetting and some key problems in implementation are solved. An improved division method is proposed to enhance the accuracy of offset curve generation in Section 3. As a new problem in 3D curve offsetting, overlapping can be eliminated by the method presented in Section 4. A break and intersection/overlapping transition method is developed in Section 5, which includes a positive weights spherical rational quartic Bezier curve generation algorithm. In Section 6, two practical engineering design cases show the practicability of the methods in this paper.

## 2 Related work

During the last 20 years, a lot of methods for the planar and geodesic curve offset approximation have been proposed and developed, which may be classified as three types: (1) Control Vertex Offset Algorithm (Coquillart 1987), i.e., moving the control vertex of NURBS curve to be offset under given precision. (2) Circle Envelope Approximation Algorithm (Lee et al. 1996). The offset is the enveloping curve by rolling a circle along the base curve. The actual offset is obtained by convolution and has degree 3p-2 for non-rational curves and 5p-4 for rational ones (Piegl and Tiller 1999), where p is the degree of the original curve. 3) Interpolation or Fitting Algorithm (Patrikalakis and Bardis 1989, Wolter and Tuohy 1992, Piegl and Tiller 1999, Lee 2003, Sun 2012), which consists of four major steps: (1) special shape identification and handling, (2) offset curve subdivision, (3) interpolation of offset points, and (4) approximation by knot removal. However, the first method cannot deal with self-intersection problems. And the second one has two major flaws (Piegl and Tiller 1999): (1) the resulted offset curve has a large number of control points; (2) the offset curve is of a high degree, e.g., the offset of a degree three rational curve is a rational curve of degree eleven. Therefore, the Interpolation or Fitting Algorithms are prevalent in researches.

To reduce the complexity, a curve is usually subdivided into line segments. Curve division algorithm is also used in curve/curve intersection (Srijuntongsiri 2011), tool path generation (Shen et al. 2010) and meshing (Laug 2010). There are two ways: (1) Parametric algorithm (Filip 1986) which determines the subdivision points number based on secondary derivate, as used in the researchers (Piegl and Tiller 1999, Sarma and Rao 2000, Sun et al. 2004). (2) Geometric algorithm which divides a curve iteratively until it satisfies the precision condition. Piegl (2002) decomposed a NURBS curve into several Bézier segments, and divided each segments with bisection until the arch rise was under the given precision. Bae (2002) subdivided the Bézier curve by applying consecutive stepwise degree reduction processes combined with adaptive subdivision - in each degree reduction step, a Bézier curve was subdivided until the approximation error from degree reduction is smaller than the corresponding step tolerance. The terminal condition of the Bézier segment division affects the precision and efficiency of the whole curve division directly. Lane and Riesenfeld (1980) proposed a termination criterion of the bisection subdivision. And then, Zhang and Wang (2003) improved the condition by some new estimate methods for the segment deviation of degree n (n=2, 3, 4) rational Bézier curves. However the above algorithms only consider the division precision to approximate a curve itself and ignore the offset accuracy of a curve with different curvatures, which is improved in this paper

The offset curve may self-intersect locally when the absolute value of the offset distance exceeds the minimum radius of curvatures in the concave regions. Also, the offset curve can self-intersect globally when the distance between two distinct points on the curve reaches a local minimum (Maekawa 1999). In general, invalid loops caused by self-intersections need be detected and eliminated. Elber and Cohen (1991) detected local self-intersections of offset curves by checking whether the tangent fields have opposite directions. Lee *et al.* (1996) applied a plane sweeping algorithm to detect all self-intersections of planar offset curves. Choi and Park (1999) removed local invalid loops from the input PS-curve before constructing a raw offset-curve, by invoking a pair-wise interference-detection (PWID) test. Elber (2003) and Seong *et*

*al.* (2006) presented a scheme to trim both local and global self-intersections of offset curves and surfaces. The scheme was based on the derivation of an analytic distance map between the original curve and its offset. By solving one bivariate polynomial equation for an offset curve, all the local and global self-intersection regions in the offset curves can be identified. The traditional method based on interference detections by searching for all contact positions is a time consuming process. Lai (2011) proposed a new algorithm called the forward locus tracing method (FLTM) which searches for all intervals split by intersections of complicated planar curves directly and transforms 2D transversal intersection problems into 1D interval identifications. Pekerman *et al.* (2008) presented several algorithms for self-intersection detection, and possible elimination, in freeform planar curves.

Compared with prior works, this paper focus on three aspects of 3D curve offset in engineering design:

1) Considering the accuracy effect by the different offset distances and curvatures, and improving the division method.

2) Eliminating the overlapping region.

3) Connecting the discontinuous or low-continuous regions automatically when offsetting $C^0/G^0$ continuous curves.

# 3 NURBS curve subdivision

## 3.1 Traditional subdivision method

In this paper, it is assumed that all the curves are described in the NURBS form. The detailed mathematical description of the NURBS can be found in literature (Sun *et al.* 2004). The main idea of NURBS curve subdivision is listed as follows (Piegl and Tiller 1999):

1) Decomposing the NURBS entity into Bézier pieces.

2) Sampling each Bézier piece uniformly at (p+1) points for non-rational, and 2(p+1) points for rational, where p is the degree.

3) At these points, computing the geometric entities such as curvatures and unit normals.

4) Subdividing the Bézier piece into line segments to be offset with the theorem 1 (Filip 1986) as following:

**Theorem** 1. Let $f(t):[a,b] \to \Re^n$ be any $\mathbf{C}^2$ curve and let $\mathbf{l}(t)$ be the linearly parameterized line segment with $\mathbf{f}(a) = \mathbf{l}(a)$ and $\mathbf{f}(b) = \mathbf{l}(b)$. Then

$$\sup_{a \le t \le b} \|\mathbf{f}(t) - \mathbf{l}(t)\| \le \tfrac{1}{8}(b-a)^2 M_0, \quad\text{where } M_0 = \sup_{a \le t \le b} \|f''(t)\| \tag{3}$$

Therefore, a certain number of division line segments are sufficiently accurate to define the curve. The number of subdivision points is given by (Sun *et al.* 2004):

$$n = \sqrt{\frac{M}{8\varepsilon}} \tag{4}$$

where $\varepsilon$ is the tolerance, $M$ is the bound on the second derivatives of the sampled points on the original curve.

The subdivision algorithm does not consider the effect of different offset distances. Suppose the original curve is divided with uniform spacing. As shown in Figure 2(A), when the original curve is a line segment, i.e., the curvatures of every point on it are 0, all the lengths of the subdivision line segments are $\Delta s$ before and after offsetting and whatever the offset distance is. When the original curve is a non-linear curve, however, the line segment lengths increase with the offset distance and the curvature radius. We take an arc for example, as shown in Figure 2(B). It is assumed that the curvature radius of every point on it is $r$ and the offset distance d is $r/2$, and the angle between two lines $OP_{0i}$ and $OP_{0i+1}$ is $\alpha$.

After offsetting, the lengths of the line segment $\Delta s^o$ between $P_{0i}$ and $P_{0i+1}$ is:

$$\Delta s^o = 2(r+d)\sin\tfrac{\alpha}{2} \qquad (5)$$

$\Delta s^o$ are varied from the curvature radii on the curve. Thus, the uniform subdivision line segments are non-uniform after offsetting when the curve has different curvatures.

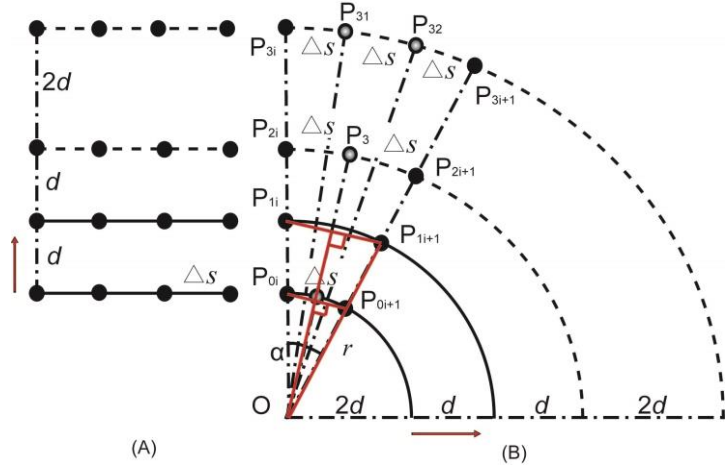

Figure 2. Offsetting of a line and an arc on different distance

However, the accuracy error of the approximation offset method based on curve interpolation or fitting depends on the max distance among interpolated or fitted points (Lucas 1974). For example, the error of the cubic spline interpolation is $O(\Delta s_{max}^4)$, where $\Delta s_{max}$ is the largest length among the line segments. Hence, when a curve is subdivided into line segments for offsetting, the accuracy effect of the offset distance should be considered here. Keeping the uniform distribution of the line segment lengths after offsetting can improve the accuracy of the offset curve generation.

3.2 **Improved subdivision method**

From Equation (5), the increase rate of a line segment length $\Delta s$ is:

$$\frac{\Delta s^o - \Delta s}{\Delta s} = \frac{2(r+d)\sin\frac{\alpha}{2} - 2r\sin\frac{\alpha}{2}}{2r\sin\frac{\alpha}{2}} = \frac{d}{r} \qquad (6)$$

where $\Delta s$ and $\Delta s^o$ denote the line segments lengths before and after offset respectively. The offset distance is $d$ and the curvature radius is $r$. When $d = r$, the increased length of the line segment is twice original one, so inserting a midpoint can

uniform the spacing distances. The density *m* of additional subdivision points is given by:

$$m = \left\lfloor \frac{d}{r} \right\rfloor \quad (7)$$

$\lfloor \ \rfloor$ denotes the floor operator. Because the curvature radii of the points on the original curve are not equal, the mean value of the curvature radii of two neighbor points generated by traditional subdivision method is used as *r* in the equation. After offsetting, the length variance of each line segment is less than $\Delta s$, i.e., the accuracy error of the offset curve generation by interpolation is determined by $\Delta s$.

An offset of a 3D cubic Bézier curve is taken as an example to show the process and effect of the improved subdivision algorithm. The Bézier curve is defined by four control points, (200,200,200), (300,500,300), (400,600,500) and (600,200,600) and the offset distance is 400. First, 20 points shown in Figure 3(A) are generated by the traditional subdivision method when the precision $\varepsilon$ is set at 1. And then, the improved method produced 83 additional subdivision points inserted in the intervals of the 20 points as shown in Figure 3 (C). For the two methods, the mean value, max value and standard deviation (SD) of the subdivision line segment lengths are shown in Table 1. By offsetting and interpolating the line segments with cubic spline, the offset curves are shown in Figure 3 (B) and (D). As shown in Table 1, the max length of the subdivision line segments for the traditional method is enlarged by 100.0%, but for the improve one, it is only 18.4%. 500 points are sampled uniformly on the offset curves to check the accuracy of the offset curve generation. The SD of the offset error by the improved method is reduced by 99.761% over the traditional one from $2.280 \times 10^{-2}$ to $5.470 \times 10^{-5}$.

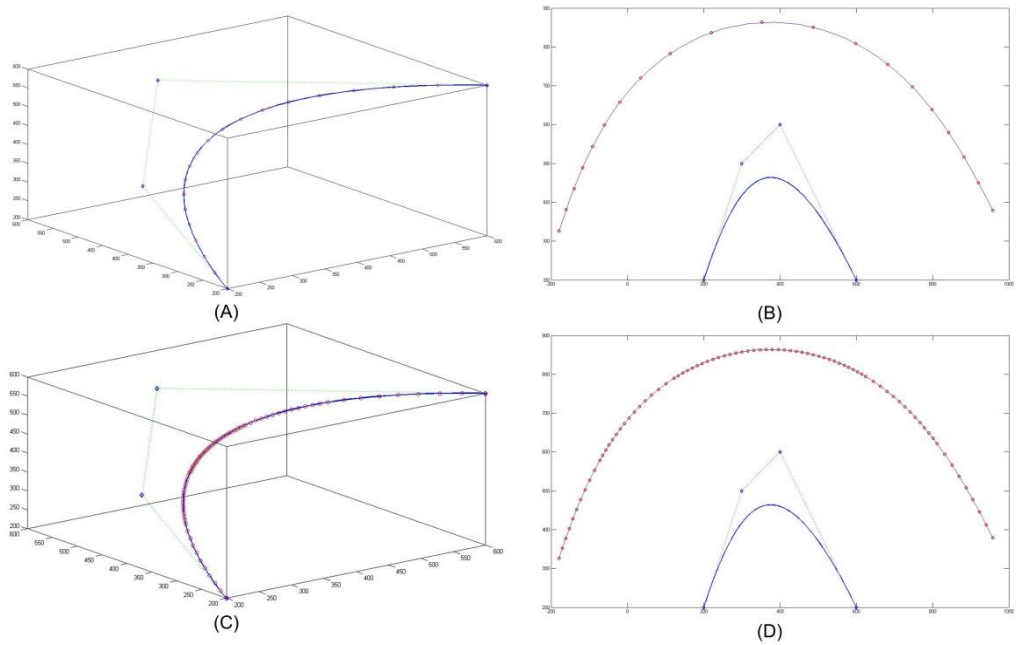

Figure 3. Subdivided and offset points by the two division methods

Table 1 Comparison of the two division methods

|  | Traditional division method | | | Improved division method | | |
| --- | --- | --- | --- | --- | --- | --- |
|  | Mean | Max | SD | Mean | Max | SD |
| Before offset | 42.822 | 68.616 | 10.666 | 9.927 | 32.720 | 8.139 |
| After offset | 88.517 | 137.555 | 25.191 | 20.535 | 38.739 | 7.132 |
| Interpolation | 399.998 | 400.030 | **$2.280 \times 10^{-2}$** | 400.000 | 400.003 | **$5.470 \times 10^{-5}$** |

## 4 Overlapping elimination

For 2D curves, the offset curves shown in Figure 4 are split into small loops by the self-intersection points. Some of these loops are invalid and must be removed; these are local invalid loops and global invalid loops. A local invalid loop is only bounded by one self-intersection point, while a global invalid loop is bounded by a pair of self-intersection points.

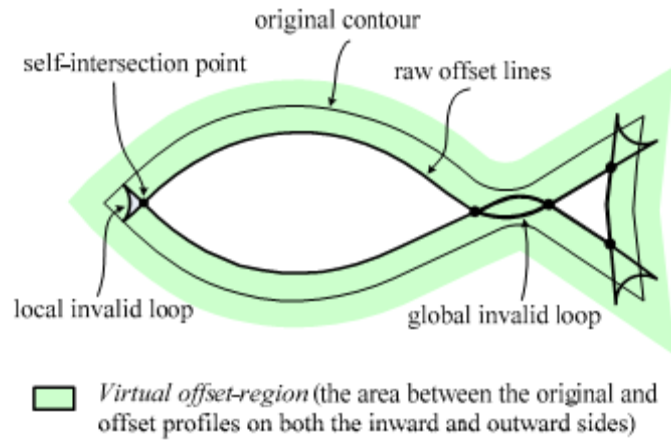

Figure 4. Raw offset curves (Lee *et al.* 2009)

For 3D curve offsetting in engineering design, another problem, overlapping, may occur. Figure 5 shows the different views of an overlapping curve. It does not self-intersect as shown in Figure 5(A), but in the parting direction (Z-axis) as shown in Figure 5(B), it makes some part of the parting surface invisible which cannot separate the mold into two parts, the core and the cavity. These problems cannot be solved by current CAD software automatically. In this section, an overlapping region elimination method is proposed to recast a valid offset curve for ruled surface generation in a straightforward way.

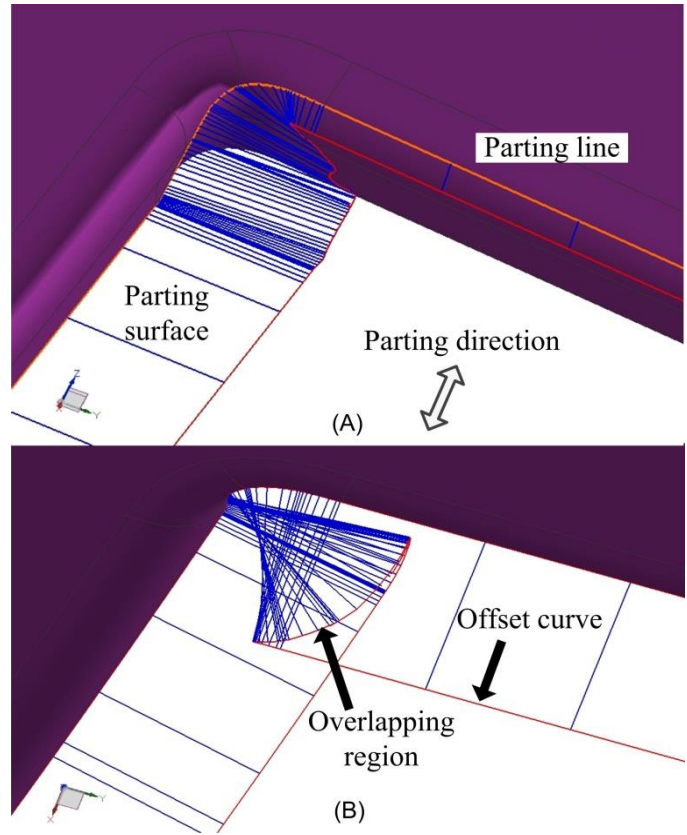

Figure 5. 3D offset curve and overlapping region on parting direction

After subdivision and offsetting, a raw offset curve is obtained, which is made up of line segments. As shown in Figure 6, the outer curve is the raw offset one from the original curve (inner one). Point A, B and C are a planar local intersection, a global intersection and an overlapping point, respectively. The major steps of the algorithm for overlapping elimination are summarized as follows:

(1) For detecting the overlapping points, the raw offset curve is projected on an arbitrary plane perpendicular to the parting direction, and then the overlapping problem becomes an intersection problem of planar curves.

(2) Each line segments is selected from the raw curve along the arrow direction. If the line segment intersects with the one selected in the sequence of the line segments, the segments between the two intersecting lines are composed of an invalid loop which will be eliminated. For example, as shown in Figure 6, the line segment $P_{i-1}$-$P_i$ intersects with $P_j$-$P_{j+1}$ at the cross point C on the project plane. The line segments between $P_i$ and $P_j$ will be removed.

(3) With the projection reverse computation, two overlapping points $P_{ic}$ and $P_{jc}$ on the raw offset curve can be found from the point C.

(4) The invalid loop (near the point B) between $P_{ic}$ and $P_{jc}$ is eliminated. And then the trimmed raw offset is obtained.

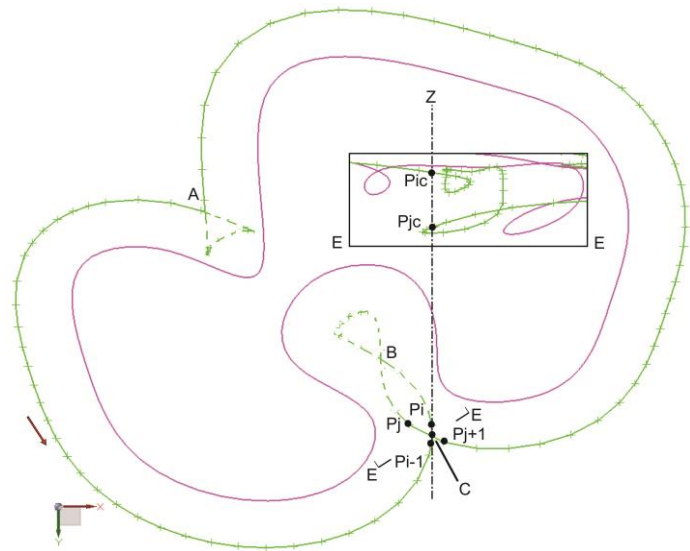

Figure 6. Illustration of overlapping elimination

The elimination approach can also detect local and global self-intersections of planar curves. After interpolation of the line segments in the trimmed raw offset curves, the final offset curves are obtained as shown in Figure 7 which contains two difference views of the curves. The offset curves are grouped by two curves, one from A to C and another form C to A, which have a discontinuous point A and a $C^0$ continuous point C.

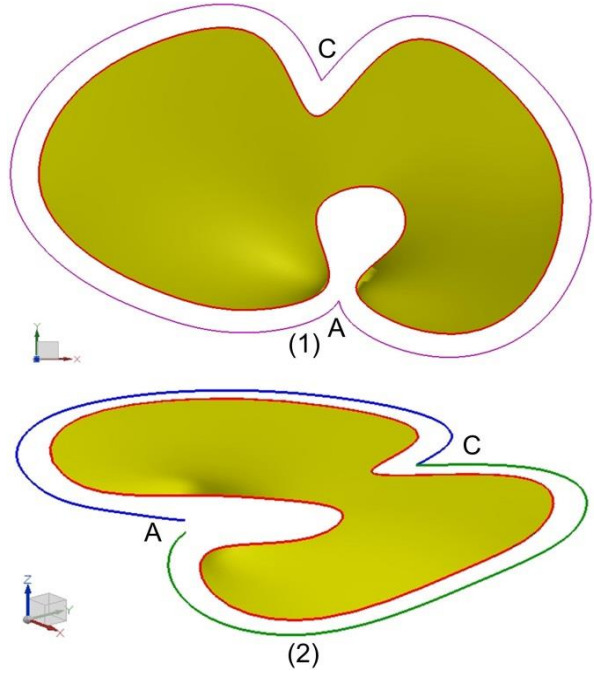

Figure 7. Offset curves after intersection/overlapping elimination

## 5 Transition

For repairing the break and intersection/overlapping problems occurred at the $C^0/G^0$ continuity points and in the discontinuous regions, a transition algorithm is proposed in this section. The discontinuous region caused by the overlapping elimination in above section can also be repaired by this method. It is assumed that $\mathbf{r}_i(t)$ and $\mathbf{r}_{i+1}(t)$ are two edges connected at P where the iso-parameter t is 1 for $\mathbf{r}_i(t)$ and 0 for $\mathbf{r}_{i+1}(t)$. After offsetting, if $\mathbf{r}'_i(1) \neq \mathbf{r}'_{i+1}(0)$, i.e. $\mathbf{r}_i(t)$ and $\mathbf{r}_{i+1}(t)$ are first-order-discontinuous, the following method will repair the offset curves. Let $\mathbf{s} = -\mathbf{r}'_i(1) \times \mathbf{r}'_{i+1}(0)$. If $\mathbf{s} \cdot \mathbf{z} < 0$, i.e. convex offset condition shown in Figure 8(a), the method in Section 5.1 attends to the situation. Otherwise it is concave offset condition shown in Figure 8(b), and the method in Section 5.2 handled.

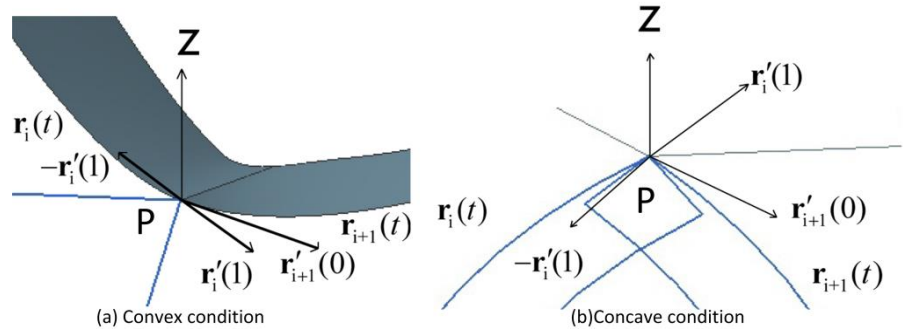

Figure 8. Two offset conditions in $C^0/G^0$ continuous curves

## 5.1 Convex offset

As shown in Figure 9, a break occurs at the co-vertex after offsetting; therefore, a transition curve is needed to bridge the gap. The general method inserts a circle or helix curve, but it cannot ensure the $C^1$ continuous connection. At the same time, the transition curve has to satisfy the requirement of equidistant offset. Spherical curves just meet the requirements. Wang and Qin (2000) proposed a method to solve a rational Bézier curve on a sphere, which attains the continuous property at two vertices. However, the curve with negative weights cannot be drawn out in most design systems, because the negative weights may bring the singularity (Yang et al. 2006). In order to construct a transition curve with the continuous and equidistant properties, a new algorithm for generating positive weights spherical rational quartic Bezier curve is proposed in this section.

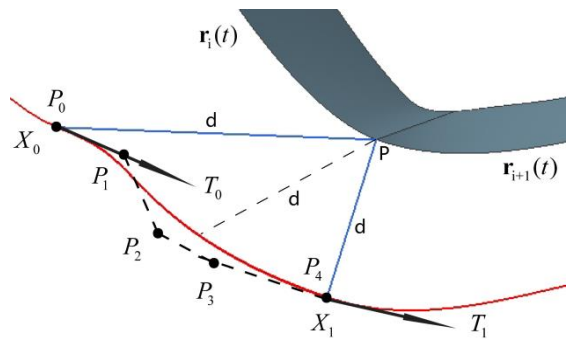

Figure 9. The transition curve in convex offset

Wang and Qin (2000) proved that, there exists a spherical quadratic rational Bézier curve that satisfies $C^1$ continuity at two boundary connection. Suppose $\mathbf{X}_0$、 $\mathbf{X}_1$、$\mathbf{T}_0$、$\mathbf{T}_1$ represent two position vectors and two tangent vectors, respectively, as shown in Figure 9. The form of a quadratic rational curve is shown as following:

$$\mathbf{p}(t) = \frac{\sum_{m=0}^{4} w_m \mathbf{P}_m B_{m,4}(t)}{\sum_{m=0}^{4} w_m B_{m,4}(t)} \quad (8)$$

where $t \in [0,1]$, $w_m \in [0,1]$, $B_{m,4}(t)$ is Bernstein basis polynomials of degree 4, Controlling point $\mathbf{P}_0 = \mathbf{X}_0$, $\mathbf{P}_1 = \mathbf{X}_0 + \frac{w_0}{4w_1}\mathbf{T}_0$, $\mathbf{P}_3 = \mathbf{X}_1 - \frac{w_4}{4w_3}\mathbf{T}_1$, $\mathbf{P}_4 = \mathbf{X}_1$.

$\mathbf{P}_2$ is supposed to be near the mid-point between $\mathbf{X}_0$ and $\mathbf{X}_1$, so that

$$\mathbf{P}_2 = \begin{pmatrix} d(2p_{2\alpha}-1)\cos(\pi(2p_{2\beta}-1))\sin(\pi(2p_{2\gamma}-1)) \\ d(2p_{2\alpha}-1)\cos(\pi(2p_{2\beta}-1))\cos(\pi(2p_{2\gamma}-1)) \\ d(2p_{2\alpha}-1)\sin(\pi(2p_{2\beta}-1)) \end{pmatrix} + \frac{(\mathbf{X}_0+\mathbf{X}_1)}{2} \quad (9)$$

where $d$ is the offset length, $p_{2\alpha}, p_{2\beta}, p_{2\gamma} \in [0,1]$, let $w_2 = 1$ for reducing unknown variable quantities.

We use optimization method to solve the eight parameters of the sphere curve. The optimization objective is taken as the mean square distance between $J$ uniform points and the vertex $P$, i.e.

$$\begin{aligned} f &= \sqrt{\frac{1}{n}\sum_{i=1}^{J}(Y_i - \bar{Y})^2} \\ Y_i &= \left|p\left(\tfrac{i}{J}\right) - P\right| \\ \bar{Y} &= \frac{1}{n}\sum_{i=1}^{J} Y_i \end{aligned} \quad (10)$$

In order to maintain $w$ positive, the method makes a punishment to the objective value once $w$ is less than 0.

Therefore, the problem of solving a positive weights spherical rational curve is transformed to a nonlinear programming problem which can be solved by many efficient approaches. This study chooses the Particle Swarm Optimization (PSO) method for its advantage over dealing with real number optimization (Wang et al. 2007). The main concept of particle swarm optimization is to minimize nonlinear functions using particle swarm methodology inspired by simulating social behavior and related to bird flocking, fish schooling and swarming theory. Each individual is treated as a volume-less particle (a point) in the N-dimensional search space.

The $i$th particle is represented as $\mathbf{U}_i = \left( w_0^i, w_1^i, w_3^i, w_4^i, p_{2\alpha}^i, p_{2\beta}^i, p_{2\gamma}^i \right)$ which consist of seven unknown parameters of the sphere curve. The best position of the $i$th particle at the $k$th iteration is recorded and represented as $\mathbf{h}_i(k)$. The index of the particle having the best value among all the particles in the populations is represented by the symbol $\mathbf{g}(k)$. The rate of the position change (velocity) for particle $i$ is represented as $\mathbf{V}_i = (V_{i1}, V_{i2}, \cdots, V_{iN})$. Then, the particles are manipulated according to the following equation at the $k+1$ iteration:

$$\mathbf{V}_i(k+1) = e\mathbf{V}_i(k) + c_1 r_1 [\mathbf{h}_i(k) - \mathbf{U}_i(k)] + c_2 r_2 [\mathbf{g}(k) - \mathbf{U}_i(k)] \qquad (11)$$

and

$$\mathbf{U}_i(k+1) = \mathbf{U}_i(k) + \mathbf{V}_i(k+1) \qquad (12)$$

where $e$ is the inertia weight in the range of [0 1], $c_1$ and $c_2$ are two positive constants represent the cognitive and social parameters, respectively, which are usually set at

2(Kennedy and Eberhart 1995), and $r_1$ and $r_2$ are two random functions in the range of [0 1].

An example is taken to show the process. As shown in Figure 9, two curves have a $C^0$ continuous co-vertex $P(0,0,0)$. The positions and tangent vectors of the break are $\mathbf{X}_0(47.553, 0, 15.451)$, $\mathbf{X}_1(29.389, 40.451, 0)$ and $\mathbf{T}_0(-3.744, 53.960, 11.524)$, $\mathbf{T}_1(-36.693, 26.659, -18.325)$, respectively. The PSO algorithm is run with a swarm size of 300, $e= 0.9$. The number $J$ in fitness function (10) is set at 100. Figure 10 shows the solution history of the result using PSO.

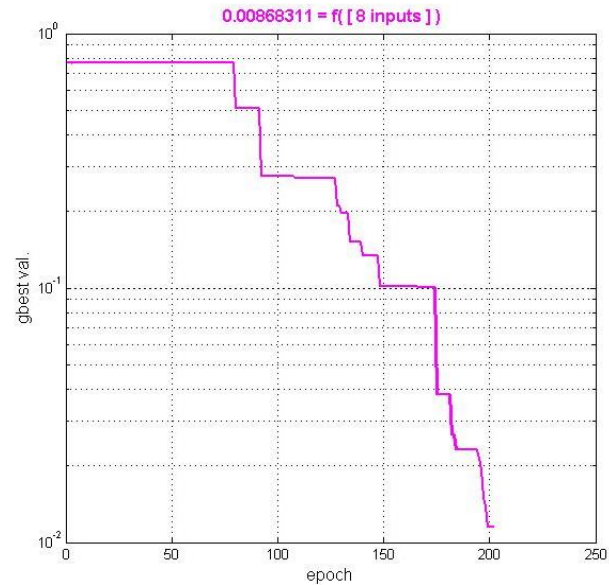

Figure 10. Solution of the PSO

After 205 iterations, the stop precision is reached, and the fitness value $g$ is 8.68e-2. Finally, $w_0 \sim w_4$ are 1.00, 0.658, 0.542, 0.585, 0.911, and $p_{2\alpha}, p_{2\beta}, p_{2\gamma}$ are 0.999, 0.989, 0.919. From equations (8) and (9), the control points obtained are $\mathbf{P}_0(47.553, 0, 15.451)$, $\mathbf{P}_1(46.617, 13.490, 18.332)$, $\mathbf{P}_2(45.045, 21.302, 10.669)$, $\mathbf{P}_3(43.669, 30.076, 7.131)$, $\mathbf{P}_4(29.389, 40.451, 0)$. With the control points and weights,

a positive weights spherical quartic Bézier can be drawn in CAD and shown in Figure 9.

## 5.2 Concave offset

At the co-vertex of the concave offset, the offset curves may intersect or overlap. In a 2D space, there is one intersection point of two offset curves. In the 3D state, in contrast, there may be no intersection point for two spatial curves, as shown in Figure 11, however, they cannot produce a high-quality continuous ruled surface. Therefore, a transition method is needed to deal with it and recast a smooth offset curve as shown in Figure 11(B).

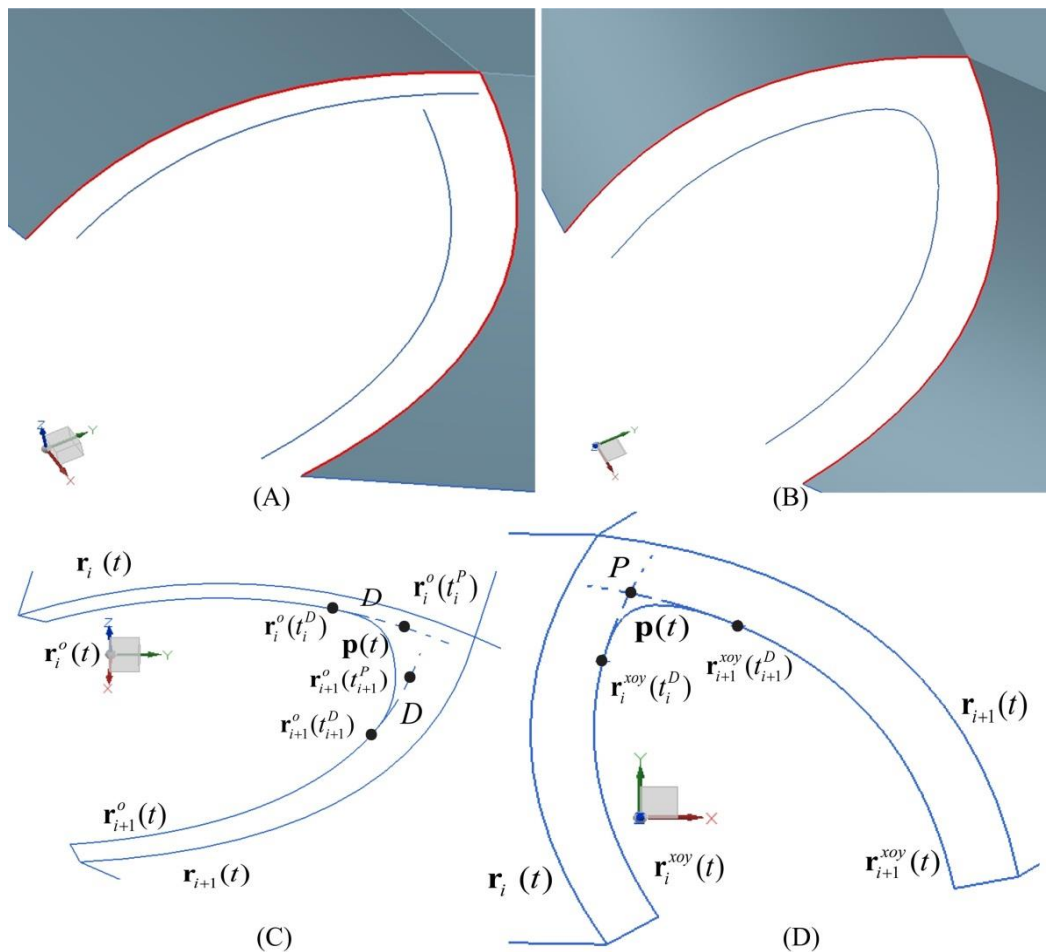

Figure 11. Concave offset and the transition curve

The transition process is list as follows:

(1) As shown in Figure 11(C), suppose two offset curves, $\mathbf{r}_i^o(t)$ and $\mathbf{r}_{i+1}^o(t)$, are $C^0$ continuous and concave connection, hence they will overlapped after offsetting, and the parting direction is oriented in the same direction as Z axis. By projecting them to an arbitrary plan perpendicular to the parting direction, e.g. XOY plane, two planar curves, $\mathbf{r}_i^{xoy}(t)$ and $\mathbf{r}_{i+1}^{xoy}(t)$, are obtained, as shown in Figure 11(D).

(2) Assuming that the curvatures of two offset curves change subtly near the overlapping region. As shown in Figure 11(D), the cross point $P$ of $\mathbf{r}_i^{xoy}(t)$ and $\mathbf{r}_{i+1}^{xoy}(t)$ can be obtained by nonlinear equation methods such as Newton iterative method, conjugate gradient method, etc. By inverse computation, the overlapping points $\mathbf{r}_i^o(t_i^P)$ and $\mathbf{r}_{i+1}^o(t_{i+1}^P)$ on $\mathbf{r}_i^o(t)$ and $\mathbf{r}_{i+1}^o(t)$, respectively, are found, as shown in Figure 11 (C).

(3) By the dichotomy method, point $\mathbf{r}_i^o(t_i^D)$ can be found on $\mathbf{r}_i^o(t)$ with a distance $D$ from the point $\mathbf{r}_i^o(t_i^P)$, as shown in Figure 11(C). The offset curve $\mathbf{r}_i^o(t)$ is trimmed by the point $\mathbf{r}_i^o(t_i^D)$. With the same way, $\mathbf{r}_{i+1}^o(t)$ is cut by the point $\mathbf{r}_{i+1}^o(t_{i+1}^D)$.

(4) Between $\mathbf{r}_i^o(t_i^D)$ and $\mathbf{r}_{i+1}^o(t_{i+1}^D)$, a $C^1$ continuous thrice Bézier curve $\mathbf{p}(t)$ is created to bridge the disconnection region. The trimming length $D$ can adjust the size of the transition region. The larger $D$, the less sharp the transition section.

# 6 Design cases

The proposed method is implemented in C# based on Siemens NX, where some basic API functions are offered to create and operate curves, such as line, arc and NURBS. Two 3D offset cases are taken as design examples which have a wide range of features, like round corners, sharp corners. All the calculating steps are executed on a PC with an Intel(R) Core(TM) i3 CPU 530 with 2.93 GHz, 4 GB of RAM, and Windows 2008 as Operating System.

## 6.1 Horizontal offset

In mold design, parting surface design is one of the most important and difficult process, which can be formed by extruding portions of the parting line along directions that are perpendicular to the parting direction of the mold insert (Li 2003). Since the parting direction is generally set at Z-axis of the coordinate, the extruding directions of the parting curves are horizontal and outward to the product model. Figure 12 shows a parting surface design case of an injection mold for a car instrumental board part, where an overflow area shown in Figure 12(B) around the 3D parting curves is modeled as a ruled surface connecting the parting curves and their offset ones which usually have break/intersection/overlapping problems. The algorithm proposed in this paper can offset the 3D parting curves by eliminating and smoothing the overlapping region automatically to create a smooth parting surface, with which the designer can generate the core and the cavity of the injection mold easily. Figure 12(C) shows the movable half of the mold.

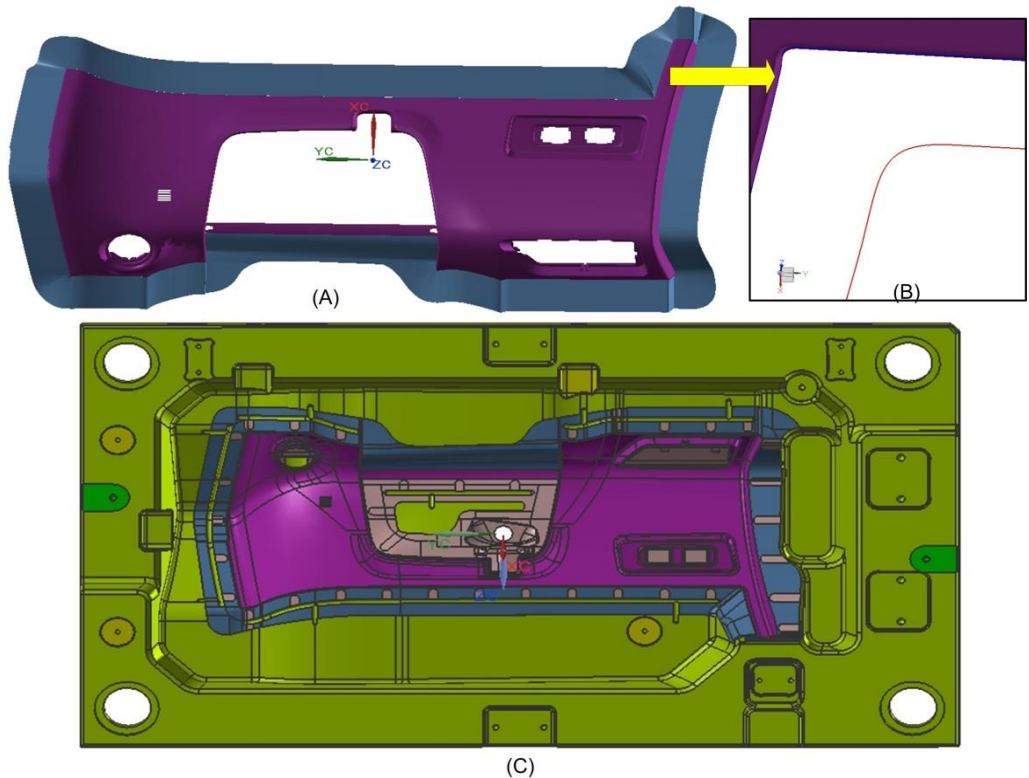

Figure 12. Parting surfaces generation and the injection mold design of an instrument board part

## 6.2 Extending offset

In checking fixture design, the checking surfaces of the checking components are usually obtained by extending the boundary of the surfaces to be checked along their tangent direction. Figure 13(A) shows a sheet metal product with a flange (checking surfaces). Note that the outer boundary curve of the flange is obtained by 3D offsetting of the inner boundary, and the flange is modeled as a ruled surface connecting the outer boundary. Because the inner boundary, i.e. the contour curve of the product, has many $C^0/G^0$ continuous points, such as $Q_1$, $Q_2$, $Q_3$ and $Q_4$, as shown in Figure 13(B), which may result in break and overlapping mentioned above. The algorithm proposed in this paper can create those ruled surfaces smoothly without manual operations. Finally, the designer may generates the checking block by stretching those ruled surfaces from the flange to the top of the workbench, as shown in Figure 13(C).

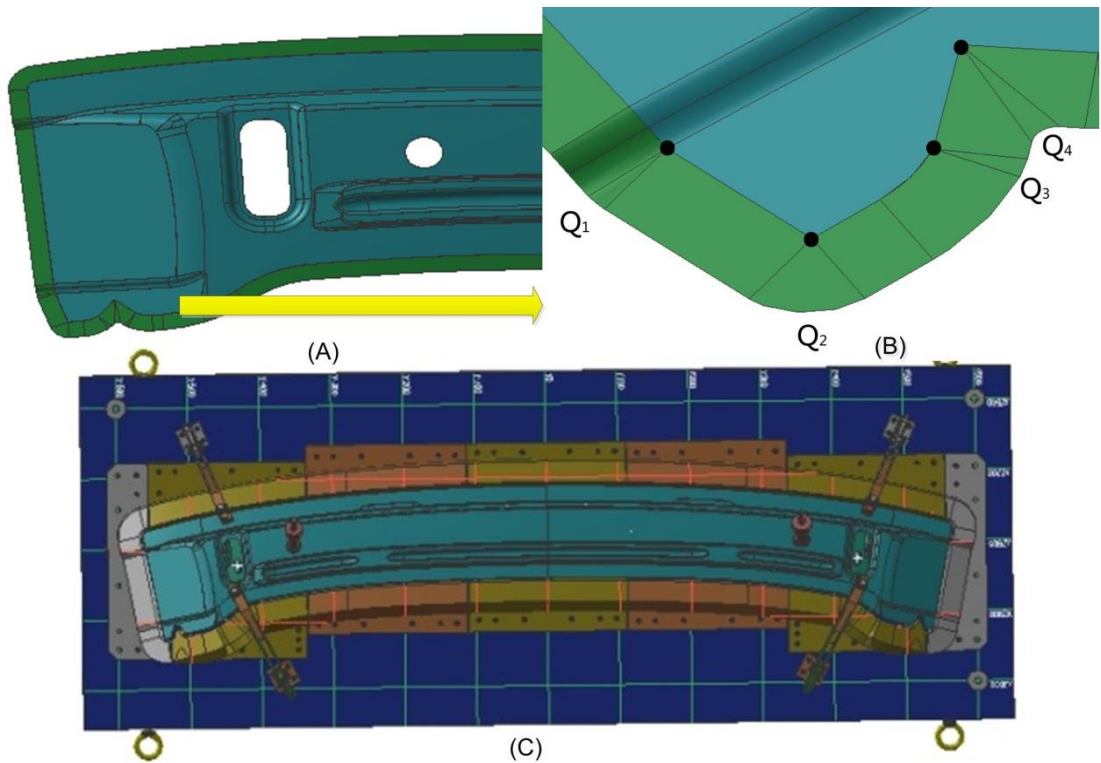

Figure 13. Checking surface generation and the checking fixture of a car fender

# 7 Conclusion

In this paper, a 3D curve offset approximation algorithm is proposed, which contains all key process solutions for ruled surface generation in engineering design, such as curve division, break/intersection/overlapping elimination and transition. For considering the accuracy effect of the different offset distances and curvatures, the improved curve division method enhances the accuracy of the offset generation. Different from 2D curves, overlapping, as a new problem in 3D offset curve, is eliminated by an overlapping detection method. To tackle some discontinuous or low-continuous regions that may result in breaking or overlapping ruled surfaces, a transition method is presented, which can bridge or smooth the regions with two approaches for convex and concave situations, respectively. In this method, a new algorithm for generating positive weights spherical rational quartic Bezier curves is proposed to bridge the breaks of offset curves.

The proposed algorithm is implemented in C# and embedded in Siemens NX CAD system. Two engineering cases, checking surface and parting surface generation, demonstrate that the approach enhances the automation level and the efficiency in engineering design.

**Acknowledgement**

Funding support from National Natural Science Foundation of China (Grant No.60903111) and Shanghai Shen Mo Die & Mold Manufacturing Co., Ltd, it is gratefully acknowledged.